\documentclass[iop,apjl]{emulateapj}
\usepackage{natbib}
\usepackage{graphicx}
\usepackage{apjfonts}
\usepackage{multirow}
\usepackage{fancyref}
\usepackage{hyperref}
\usepackage{xcolor}
\hypersetup{
	colorlinks,
	linkcolor={red!50!black},
	citecolor={blue!75!magenta},
	urlcolor={blue!80!black}
}

\shorttitle{High-J CO IR Correlations}
\shortauthors{D. Liu et al.}

\begin{document}


\title{High-{\it{J}} CO Versus Far-Infrared Relations in Normal and Starburst Galaxies}


\author{
		Daizhong Liu		\altaffilmark{1,2,3},
		Yu Gao				\altaffilmark{1},
		Kate Isaak			\altaffilmark{4},
		Emanuele Daddi		\altaffilmark{3},
		Chentao Yang		\altaffilmark{1,2,5},
		Nanyao Lu			\altaffilmark{6},
		Paul van der Werf	\altaffilmark{7} 
}

\altaffiltext{1}{Purple Mountain Observatory / Key Laboratory for Radio Astronomy, Chinese Academy of Sciences, Nanjing, China; yugao@pmo.ac.cn}
\altaffiltext{2}{University of Chinese Academy of Sciences, Beijing, China}
\altaffiltext{3}{CEA Saclay, Laboratoire AIM-CNRS-Universit\'{e} Paris Diderot, IRFU / SAp, Orme des Merisiers, 91191 Gif-sur-Yvette, France}
\altaffiltext{4}{Scientific Support Office, ESTEC/SRE-S Keplerlaan 1, NL-2201 AZ, Noordwijk, The Netherlands}
\altaffiltext{5}{Institut d'Astrophysique Spatiale, B\^{a}t 121, Universit\'{e} Paris-Sud, 91405 Orsay, France} 
\altaffiltext{6}{Infrared Processing and Analysis Center, California Institute of Technology, MS 100-22, Pasadena, CA 91125, USA}
\altaffiltext{7}{Leiden Observatory, Leiden University, Post Office Box 9513, NL-2300 RA Leiden, The Netherlands}




\begin{abstract}

We present correlations between 9 CO transitions ($J=4-3$ to $12-11$) and beam-matched far-infrared (far-IR) luminosities ($L_{\mathrm{FIR},\,b}$) among 167 local galaxies, using {\it{Herschel}} Spectral and Photometric Imaging ReceiverFourier Transform Spectrometer (SPIRE; FTS) spectroscopic data and Photoconductor Array Camera and Spectrometer (PACS) photometry data. We adopt entire-galaxy FIR luminosities ($L_{\mathrm{FIR},\,e}$) from the {\it{IRAS}} Revised Bright Galaxy Sample and correct to $L_{\mathrm{FIR},\,b}$ using PACS images to match the varying FTS beam sizes. All 9 correlations between $L'_{\mathrm{CO}}$ and $L_{\mathrm{FIR},\,b}$ are essentially linear and tight ($\sigma=0.2-0.3\,dex$ dispersion), even for the highest transition, $J=12-11$. This supports the notion that the star formation rate (SFR) is linearly correlated with the dense molecular gas ($n_{\mathrm{H}_2}\gtrsim10^{4-6}\,cm^{-3}$). We divide the entire sample into three subsamples and find that smaller sample sizes can induce large differences in the correlation slopes. We also derive an average CO spectral line energy distribution (SLED) for the entire sample and discuss the implied average molecular gas properties for these local galaxies. We further extend our sample to high-{\it{z}} galaxies with literature CO($J=5-4$) data from the literature as an example, including submillimeter galaxies (SMGs) and "normal" star-forming BzKs. BzKs have similar FIR/CO(5--4) ratios as that of local galaxies, and follow well the locally-determined correlation, whereas SMG ratios fall around or slightly above the local correlation with large uncertainties. Finally, by including Galactic CO($J=10-9$) data as well as very limited high-{\it{z}} CO($J=10-9$) data, we verify that the CO($J=10-9$)--FIR correlation successfully extends to Galactic young stellar objects, suggesting that linear correlations are valid over 15 orders of magnitude. 
	
\end{abstract}

\keywords{galaxies: ISM --- galaxies: starburst --- infrared: ISM --- ISM: molecules}

\bibliographystyle{apj}

\defcitealias{Carilli2013}{CW13}
\defcitealias{Greve2014}{G14}
\defcitealias{Bayet2009}{Bayet et al. 2009}
\defcitealias{Bothwell2013}{Bothwell et al. 2013}
\defcitealias{Yang2013}{Yang et al. 2013}
\defcitealias{Zhang2014}{Zhang et al. 2014}
\section{Introduction}

Molecular gas is the raw material from which stars are formed, hence its presence and physical conditions correlate with the star formation rate (SFR). A well-known correlation between gas surface density and SFR surface density is the {\it Kennicutt-Schmidt law} (K-S law): $\Sigma_{SFR} = A\,\Sigma_{gas}^{\;N}$, with slope $N=1.4\pm0.15$ \citep{Kennicutt1998}, where $\Sigma_{SFR}$ is based on $\mathrm{H}_{\alpha}$ luminosity or far-infrared (far-IR) luminosity ($L_{FIR}$, $40-400\mu{m}$) and $\Sigma_{gas}$ is the sum of the atomic gas (from $\mathrm{H}\mathrm{I}$ observations) and molecular gas (inferred from CO(1--0) line luminosity) surface density. This quantitative correlation serves as a fundamental input to most cosmological simulations \citep[e.g.][]{Springel2003,MacLow2004,Krumholz2005,Krumholz2007,Narayanan2008b}. However, it brushes over and/or oversimplifies the relationship between the individual gas components and star formation (SF). First, observations have shown that the presence and state of the atomic gas component has little impact on SF \citep[e.g.][]{Bigiel2008,Schruba2013,LiuLJ2012,LiuLJ2015}. Second the correlation between the most widely used molecular gas tracer, CO(1--0), and $L_{FIR}$ is found to be more complex than just a single power law: \cite{Gao2004} showed that the CO(1--0)--FIR slope changes from 1.27 to 1.73 with different sample selections, e.g. including only normal star-forming galaxies (SFGs:$\;L_{\mathrm{IR}}<10^{11}\,L_{\odot}$), or adding (ultra-)luminous IR galaxies (LIRGs:$\;L_{\mathrm{IR}}=10^{11-12}\,L_{\odot}$, ULIRGs:$\;L_{\mathrm{IR}}>10^{13}\,L_{\odot}$), respectively. \cite{Greve2014} showed similar slope variations from 0.9 to 1.4 by analyzing a large number of data from the literature. It is thus questionable that a single power-law relation exists between CO(1--0) and FIR and whether it can be applied to all galaxies even if it exists. 

CO(1--0) has a low critical density ($n_{\mathrm{H}_2,crit}\sim2\times10^{3}\,cm^{-3}$), and thus traces the total amount of molecular gas. It provides no physical insight into the higher density gas ($n_{\mathrm{H}_2}>10^{4}\,cm^{-3}$) that is known to be found at the formation sites of individual stars \citep{Kennicutt2012}. The sensitivity of the CO $(J_{u}{\to}J_{u}-1)$ transition to denser/warmer gas increases with an increasing upper level ($J_{u}$): for example, CO(10-9) has a critical density of $n_{\mathrm{H}_2,crit}\sim10^{6}\,cm^{-3}$ \citep{Carilli2013}. In contrast, the ground-{\it{J}} of high dipole-moment molecules such as HCN and HCO$^+$ already have $n_{\mathrm{H}_2,crit}\sim10^{6}\,cm^{-3}$, thus probing the dense gas in cold phase. Studies based on HCN \citep{Gao2004,Wu2005,Wu2010} and high-{\it{J}} CS \citep{Zhang2014} have found a unified linear correlation between dense gas and the SFR, describing the scenario in which all dense gas above a certain density threshold has a similar efficiency/timescale to collapse and form new stars, thus is linearly determining SFRs. 

Until recently, published mid-to-high-{\it{J}} ($J_{u}\ge4$) CO data in local galaxies were still scarce due to the challenges of observing at frequencies $\gtrsim450\,\mathrm{GHz}$ from the ground. \cite{Bayet2009} performed large velocity gradient (LVG) modeling with ground-based $J_{u}\le7$ CO to extrapolate $J_{u}>7$ transitions for several local galaxies. They found that CO versus total IR luminosity ($L_{TIR}$, $8-1000{\mu}m$) correlations have decreasing slopes for increasing $J_{u}$, similar to what modelings and simulations predicted \citep{Narayanan2008b,Juneau2009}. 

More recently, \citetalias{Greve2014} presented the {\it{Herschel}} Spectral and Photometric Imaging Receiver \citep[SPIRE;][]{Griffin2010} Fourier Transform Spectrometer \citep[FTS,][]{Naylor2010} high-{\it{J}} CO data from the HerCULES program \citep{vanderWerf2010,Rosenberg2015}. They find that up to CO(7--6) the CO versus TIR or FIR relations are all roughly linear, while for higher-$J_{u}$ the correlations become sub-linear, similar to the results of \cite{Bayet2009}. 

In this work, we use a large SPIRE FTS data set to statistically determine the correlations between 9 $L'_{\mathrm{CO}}$ ($J=4-3\;\mathrm{to}\;12-11$) and beam-matched $L_{\mathrm{FIR},b}$, and to determine which transitions are the best tracers of star formation. We also verify the slope variations against $J_{u}$ and test the validity of dense gas versus SF relation at high-{\it{z}}. The sample, data, and method are described in Section \ref{Section2}. The results and a discussion are given in Section \ref{Section3}. We adopt $H_0=73\,km\,s^{-1},\;\Omega_{\Lambda}=0.73,\;\Omega_{M}=0.27$.

\section{Sample and Data}
\label{Section2}

The sample was selected from all public FTS observations in the {\it{Herschel}} Science Archive (HSA) of local galaxies that are in the {\it{IRAS}} Revised Bright Galaxy Sample \citep{Sanders2003}. A further selection criterion requiring the availability of $70-160\,{\mu}m$ band imaging data taken with Photoconductor Array Camera and Spectrometer \citep[PACS;][]{Poglitsch2010} for galaxies partially resolved by the FTS beam (see Section \ref{Section21}) was imposed to enable beam-matching techniques to be used in the analysis. The final sample contains 167 local galaxies ($z<0.064$, $d_L<286\,\mathrm{Mpc}$), including 124 (U)LIRGs and 43 SFGs (see \cite{Rosenberg2015} and \cite{Lu2014} for details of the HerCULES and GOALS programs, respectively; details of the full data set are given in Liu et al. in preparation).

\subsection{FTS CO Measurements}
\label{Section21}

FTS has two bolometer arrays: the SLW bolometer array ($446-989\,\mathrm{GHz}$) and the SSW bolometer array: ($959-1543\,\mathrm{GHz}$). Two observing modes are used: single-pointing for point-like sources (mostly with angular size less than FTS beam sizes, and $d_L>30\,Mpc$), and mapping for nearby extended sources ($d_L<30\,Mpc$). In single-pointing observations the central bolometer of each array is  coaligned with the target, while off-axis bolometers point to the off-center sky. 
Mapping observations perform jiggling to scan extended targets, and each individual bolometer will produce one spectrum at its own pointing (R.A., decl.). All mapping data and $\sim30\%$ single-pointing data were reduced using SPIRE v12 calibration products and {\it{Herschel}} Interactive Processing Environment \citep[HIPE v12.1.0,][]{Ott2010} pipelines, with the remainder reduced using SPIRE v10 calibration products+pipelines. We note that measured line fluxes for the two sets of the $\sim30\%$ single-pointing data show little difference ($\sigma\sim8\%$). For single-pointing data, we use central bolometers to extract CO lines. 
For mapping data, which are usually for spatially-extended galaxies, we select individual bolometers that are $>$half-beam-separated (Nyquist sampling) to extract CO lines. 

Note that FTS covers a wide range of frequencies, and its beam size varies accordingly \footnote{Figure 5.18 of http://herschel.esac.esa.int/Docs/SPIRE/spire\_handbook.pdf}: $\sim43''$ at CO(4--3) to $\sim17''$ at CO(13--12), and is not a simple function of frequency. The beam size variation is well-calibrated for each central bolometer, but unavailable for off-axis bolometers. Thus we assign an additional 15\% uncertainty for off-axis bolometers. 

To derive CO line flux, we use two different line profile fitting functions in HIPE: a Sinc and a Sinc-convolved-Gaussian (SCG). A SCG, where a Gaussian line profile is convolved with the Sinc instrument response, is appropriate for observations of sources with broad/resolved lines \citep[e.g.][]{Zhao2013}: in such cases a Sinc fit would underestimate the line flux by $\sim40\%$. In the more common case where the CO line is not obviously resolved, the SCG-derived fluxes are systematically larger by $<20\%$. Given that a large fraction of single-pointing data have broad/resolved lines while mapping data do not, here we use SCG and Sinc for single-pointing and mapping data, respectively, although we note that using only SCG or Sinc for all data has no significant change to our correlation results. The CO flux uncertainties are derived from the rms in the baseline-subtracted spectra in the vicinity of each CO line rather than from formal line fitting uncertainties. 
Then the fluxes are converted to $L'_{\mathrm{CO}}$ according to \cite{Solomon1992}. 

\subsection{Beam-matched FIR}
\label{Section22}

As a starting point we took the entire-galaxy FIR luminosity ($L_{\mathrm{FIR},e}$) from \cite{Sanders2003}. Given the mismatch between the dramatically varying FTS CO beam sizes and the physical extent of nearby galaxies \citep[e.g.][]{Galametz2013} and some merging/interacting (U)LIRGs \citep[e.g.][]{Gao1999}, we used a beam-scaling method based on the PACS imaging photometry \footnote{All PACS data are updated to calibration 48 and post-processed with {\it Scanamorphos v19.0} \citep{Roussel2013}.} to scale entire-galaxy $L_{\mathrm{FIR},e}$ down to the local region that matches the CO beam size area ($L_{\mathrm{FIR},b}$, for each CO line of each bolometer). PACS bands span the peak of the spectral energy distribution (SED) of typical FIR-luminous local galaxies, and thus provide a good proxy for FIR and are insensitive to dust temperatures ($T_{\mathrm{dust}}$) \footnote{PACS bands: 70, 100, and 160${\mu}m$. $\lambda_{peak}\approx{290\,{\mu}m}/{T_{dust}}$, where $T_{\mathrm{dust}}\sim20-30\,K$ for typical local SFGs.}. By performing photometry with an aperture of CO beam size ($F_{\mathrm{PACS},b}$) and an aperture of entire-galaxy ($F_{\mathrm{PACS},e}$), respectively, we can determine the FIR luminosity that is appropriate to each CO beam size: $L_{\mathrm{FIR},b}={F_{\mathrm{PACS},b}}/{F_{\mathrm{PACS},e}}{\times}L_{\mathrm{FIR},e}$.


In this way we calculate a value of $L_{\mathrm{FIR},b}$ for each CO line. This is essential for local galaxies that are partially resolved for FTS CO beam sizes \citepalias[e.g.][]{Greve2014,Bayet2009,Yang2013,Zhang2014}. 

\section{Results and Discussion}
\label{Section3}

\begin{figure*}
	\epsscale{1.2}
	\plotone{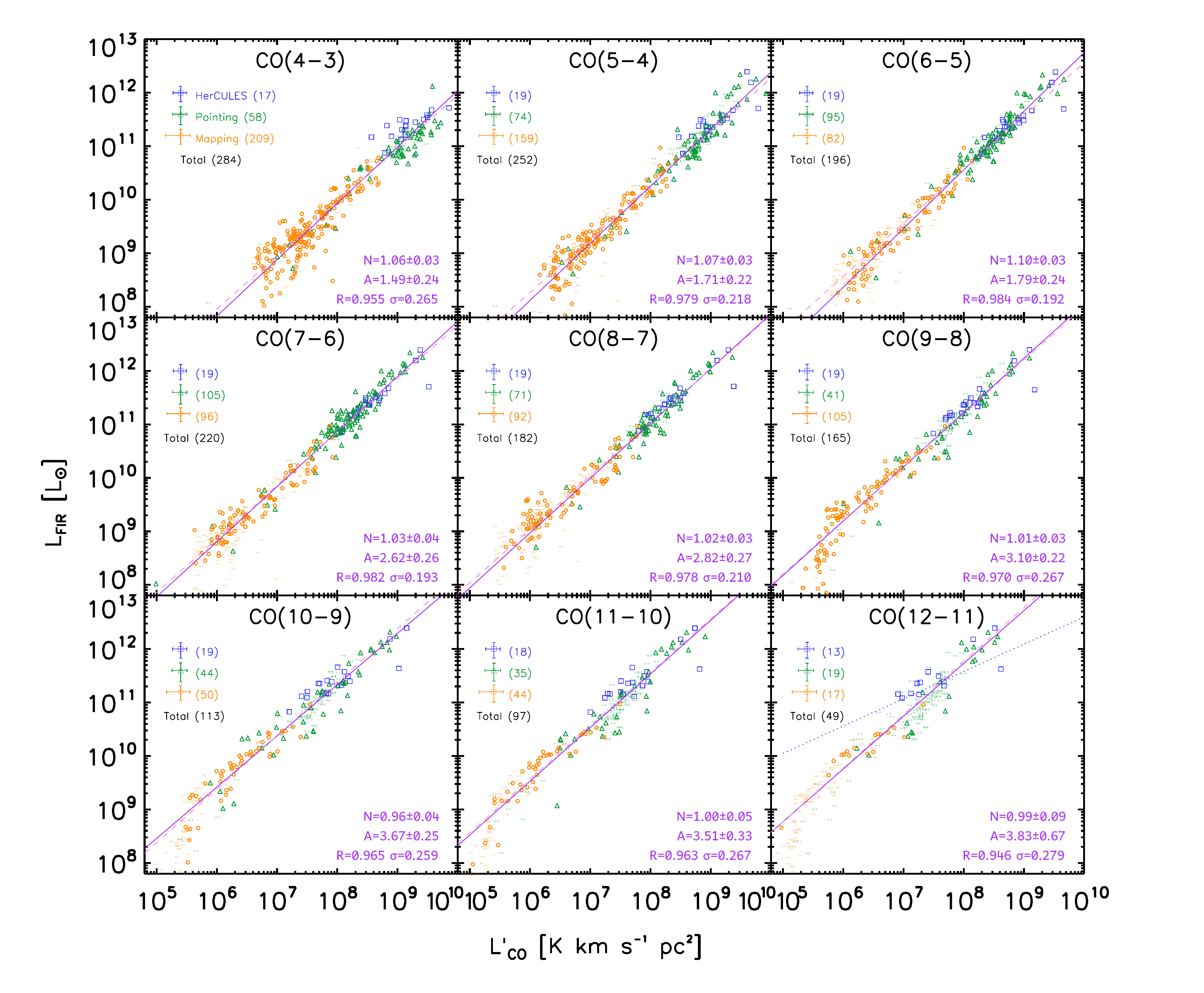}
	\caption{
Nine correlations between $L'_{\mathrm{CO}}$ and $L_{\mathrm{FIR},b}$ among local galaxies (blue squares and green triangles) and spatially resolved regions 
of nearby galaxies (orange circles). 
The blue squares are from the HerCULES sample, and are the brightest (U)LIRGs. 
The green triangles are the other single-pointing data which are less bright, consisting of (U)LIRGs, nearby Seyfert nuclei and spiral nuclei. 
The orange circles are the mapping data, of which many are nearby galaxy nuclear regions, but still have some off-nuclear regions at the highest transitions.
The numbers of detections for each (sub)sample  are shown at top left, next to the symbol legends. 
The error bar of each symbol legend is the mean uncertainties of the (sub)sample. 
The arrows are $3\sigma$ upper limits. 
The best-fitting parameters: slope $N$, intercept $A$, Pearson correlation coefficient $R$ and scatter $\sigma\;(dex)$ are shown at bottom right. 
The solid lines are the best-fit lines with free slopes, while dashed lines are the best-fit lines with slopes fixed to one, which represent the average $L_{\mathrm{FIR}}/L'_{\mathrm{CO}}$ for each transition. 
In the last panel CO(12--11) we show the best-fit line of \citetalias{Greve2014} as a blue dotted line for comparison. 
\\
\label{Fig01}}
\end{figure*}

\subsection{Correlations between $L'_{\mathrm{CO}}$ and Beam-matched $L_{\mathrm{FIR}}$}
\label{Section31}

Fig.\ref{Fig01} shows the 9 CO--FIR correlations of local galaxies and spatially resolved regions of nearby galaxies. 
FTS can observe CO(4--3) to CO(13--12) except for galaxies at $z>0.032$ (mostly (U)LIRGs) whose CO(4--3) shift out of the FTS SLW waveband. 
Moreover, detections of CO(13--12) are sparse, and thus are not analyzed here. 

To derive the slope and intercept, we fit all data points (excluding upper limits) with two linear fitting codes: an IDL least-squares fitting code based on MPFIT \citep{Markwardt2009}, and an IDL Bayesian regression code LINMIX\_ERR \citep{Kelly2007}. 
The two methods give consistent results, thus we only show the latter results in Fig.\ref{Fig01}. 

CO(4--3) has the largest number of detections, but the least dispersions are seen in the CO(6--5) and CO(7--6) panels (both with $\sigma=0.19\,dex$). 
This confirms the conclusion of \cite{Lu2014} that CO $J_{u}\sim6-7$ transitions are the best SFR tracers for (U)LIRGs. 

For the entire sample, we find no evidence of significant decrease in slopes with increasing $J_{u}$: $N\sim$1.0--1.1 at $J_{u}\sim4-6$, and 0.96--1.0 at $J_{u}\sim10-12$. 
Besides, the fittings are consistent with the average distribution of upper limits for non-detections in each transition. 

\cite{Bayet2009} analyzed the IR beam corrections based on 850$\mu{m}$ images and found correction factors similar to ours for galaxies we had in common. They found that IR beam corrections have only minor effects on the slopes ($<5\%$) and thus did not apply them, while in contrast we found that slopes without IR beam corrections are smaller by 
10--15\%. 
This partially explains the discrepancy between our results and their decreasing slopes, i.e. $N\sim1$ at $J_{u}\sim3$, then $N\sim0.8$ at $J_{u}\sim7$, and finally $N=0.53\pm0.07$ at $J_{u}=12$. 

In addition, small number statistics is another key factor behind the discrepancies between different studies. 
\citetalias{Greve2014} presented the high-{\it{J}} CO--FIR correlation with HerCULES FTS data \citep{vanderWerf2010} and data the from literature of high-$z$ submillimeter galaxies (SMGs). 
Their HerCULES sample contains 23 (U)LIRGs (excluding extended/merging galaxies). 
They evaluate IR beam correction with 870$\mu{m}$ or 350$\mu{m}$ maps, but the correction factors are small for their local sample (mostly (U)LIRGs). 
They found correlation slopes $N\sim1.0$ for $1\le{J_{u}}\le5$, which then decrease to $N=0.87\pm0.05$ at $J_{u}=7$ and then rapidly become $N=0.51\pm0.11$ at $J_{u}=12$. 

\begin{deluxetable*}{cccccccccccc}
\tablecaption{\label{Tab01}Best-fit parameters of Beam-matched CO--FIR correlations}\tablenum{1}
\tablehead{\colhead{(Sub)sample} & \colhead{CO(4--3)} & \colhead{CO(5--4)} & \colhead{CO(6--5)} & \colhead{CO(7--6)} & \colhead{CO(8--7)} & \colhead{CO(9--8)} & \colhead{CO(10--9)} & \colhead{CO(11--10)} & \colhead{CO(12-11)}}
\startdata
Entire &     ${N}=$$1.06\pm0.03$ &  $1.07\pm0.03$ &  $1.10\pm0.03$ &  $1.03\pm0.04$ &  $1.02\pm0.03$ &  $1.01\pm0.03$ &  $0.96\pm0.04$ &  $1.00\pm0.05$ &  $0.99\pm0.09$ \\
&     ${A}=$$1.49\pm0.24$ &  $1.71\pm0.22$ &  $1.79\pm0.24$ &  $2.62\pm0.26$ &  $2.82\pm0.27$ &  $3.10\pm0.22$ &  $3.67\pm0.25$ &  $3.51\pm0.33$ &  $3.83\pm0.67$ \\
& $\bar{A}=$$1.96\pm0.07$ &  $2.27\pm0.07$ &  $2.56\pm0.08$ &  $2.86\pm0.07$ &  $3.04\pm0.08$ &  $3.20\pm0.09$ &  $3.38\pm0.10$ &  $3.56\pm0.11$ &  $3.77\pm0.15$ \\
\hline
HerCULES  &     ${N}=$$0.75\pm0.62$ &  $1.08\pm0.43$ &  $0.98\pm0.29$ &  $0.84\pm0.35$ &  $0.81\pm0.24$ &  $0.81\pm0.26$ &  $0.73\pm0.25$ &  $0.76\pm0.26$ &  $0.65\pm0.31$ \\
&     ${A}=$$4.46\pm0.57$ &  $1.67\pm3.89$ &  $2.78\pm2.61$ &  $4.18\pm3.06$ &  $4.68\pm2.04$ &  $4.80\pm0.21$ &  $5.56\pm0.02$ &  $5.49\pm0.02$ &  $6.58\pm0.00$ \\
& $\bar{A}=$$2.20\pm0.23$ &  $2.40\pm0.21$ &  $2.61\pm0.21$ &  $2.85\pm0.21$ &  $3.08\pm0.21$ &  $3.28\pm0.22$ &  $3.42\pm0.22$ &  $3.65\pm0.23$ &  $3.90\pm0.28$ \\
Pointing  &     ${N}=$$1.00\pm0.10$ &  $1.10\pm0.09$ &  $1.06\pm0.09$ &  $1.05\pm0.08$ &  $1.09\pm0.11$ &  $1.02\pm0.11$ &  $0.93\pm0.09$ &  $1.05\pm0.12$ &  $1.12\pm0.18$ \\
&     ${A}=$$1.96\pm0.88$ &  $1.43\pm0.79$ &  $2.09\pm0.79$ &  $2.51\pm0.65$ &  $2.34\pm0.92$ &  $3.02\pm0.84$ &  $3.88\pm0.70$ &  $3.12\pm0.91$ &  $2.73\pm1.36$ \\
& $\bar{A}=$$1.93\pm0.13$ &  $2.30\pm0.12$ &  $2.62\pm0.11$ &  $2.89\pm0.11$ &  $3.02\pm0.13$ &  $3.18\pm0.16$ &  $3.34\pm0.16$ &  $3.51\pm0.18$ &  $3.65\pm0.24$ \\
Mapping  &     ${N}=$$1.17\pm0.10$ &  $1.13\pm0.09$ &  $1.11\pm0.10$ &  $0.96\pm0.08$ &  $0.94\pm0.08$ &  $1.17\pm0.12$ &  $1.12\pm0.18$ &  $1.03\pm0.17$ &  $0.66\pm0.42$ \\
&     ${A}=$$0.64\pm0.71$ &  $1.29\pm0.67$ &  $1.78\pm0.74$ &  $3.08\pm0.56$ &  $3.44\pm0.56$ &  $2.14\pm0.77$ &  $2.68\pm1.14$ &  $3.40\pm1.04$ &  $5.99\pm0.00$ \\
& $\bar{A}=$$1.93\pm0.08$ &  $2.22\pm0.10$ &  $2.47\pm0.13$ &  $2.82\pm0.12$ &  $3.04\pm0.12$ &  $3.19\pm0.12$ &  $3.40\pm0.17$ &  $3.58\pm0.18$ &  $3.81\pm0.29$ 
\enddata
\tablecomments{The best-fit CO -- FIR correlations for the 9 transitions. We fit the entire sample and three subsamples separately: \citetalias{Greve2014} HerCULES data, single-pointing data, and mapping data. Both $x$ and $y$ errors are considered. Upper limits are not considered in the fit. $N$ is the best-fit slope and $A$ is the best-fit intercept from free-slope fitting. $\bar{A}$ is the mean FIR/CO ratio ($\bar{A}=\log\,(L_{\mathrm{FIR}}/L'_{\mathrm{CO}})$). 
}
\end{deluxetable*}

For comparison, we divide our entire sample into three subsamples: a HerCULES subsample corresponding to the \citetalias{Greve2014} local sample (mostly (U)LIRGs, excluding extended/merging ones), all other single-pointing data (mixed SFGs+(U)LIRGs), and all mapping data (nearby resolved SFGs, dominating the faint-end). 
Table.\ref{Tab01} lists the best-fit slope $N$, the intercept $A$, and the mean FIR/CO ($\bar{A}=\log\,(L_{\mathrm{FIR}}/L'_{\mathrm{CO}})$) for the entire sample and three subsamples. 
For the HerCULES subsample, similar to \citetalias{Greve2014} and \cite{Bayet2009}, we find decreasing slopes, e.g. $N=0.65\pm0.31$ at CO(12--11). 
We overlay the significantly sub-linear CO(12-11)--FIR correlation of \citetalias{Greve2014} ($N=0.51\pm0.11$) in the CO(12--11) panel of Fig.\ref{Fig01} for reference. 

Thus, 
with the largest high-{\it{J}} CO data set available to date, we 
conclude that within the luminosity range shown in Fig.\ref{Fig01}, $L'_{\mathrm{CO}}$ $7\le{J}_{u}\le9$ closely and linearly follow $L_{\mathrm{FIR}}$ and by extension the SFRs, while all other $4\le{J}_{u}\le12$ CO--FIR correlations are tight and not far from linear. 
These results are not in conflict with previous studies.

\subsection{Average CO/FIR Spectral Line Energy Distribution (SLED)}
\label{Section32}

\begin{figure}
\centering
\includegraphics[trim=140 0 160 0, clip, width=0.9\columnwidth]{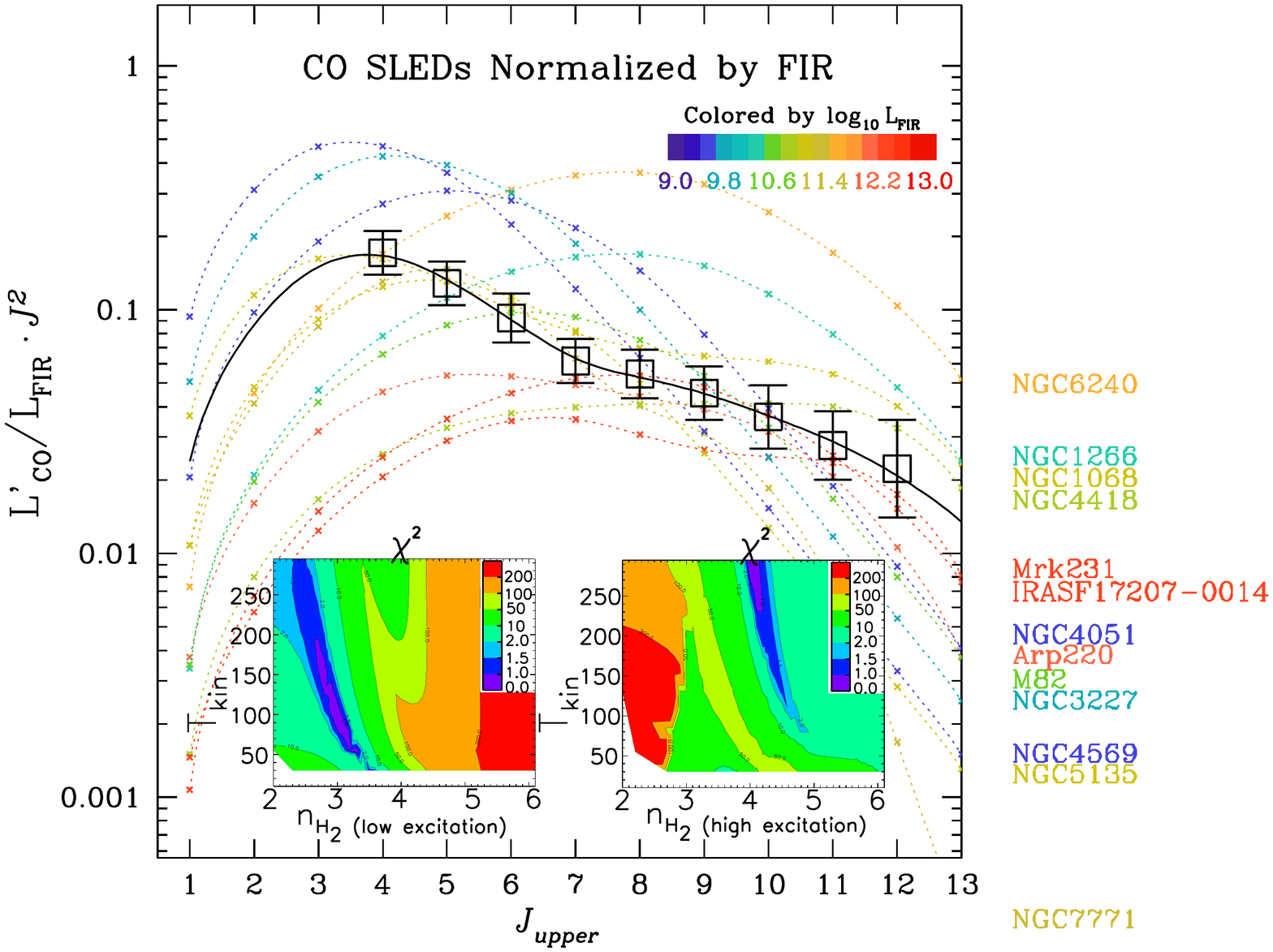}%
\caption{Average CO/FIR SLED (i.e. CO SLED normalized by FIR) for the entire sample, overlaid with several FIR-normalized CO SLEDs of individual galaxies spanning a wide range of $L_{\mathrm{FIR}}$. 
The black open squares are the values of the average CO/FIR (i.e. $L'_{\mathrm{CO}}/L_{\mathrm{FIR}} \times J_{u}^2$). 
The black solid curve is the best-fit two-component LVG model. 
The other dashed curves are the best-fit one- or two-component LVG models for individual galaxies. The colors indicate their $\lg\,(L_{\mathrm{FIR}})$. 
The two embedded panels are the least-$\chi^2$ fitting results of low-excitation gas (left) and high-excitation gas (right) respectively. 
\label{Fig04}}
\end{figure}

In Fig.\ref{Fig04}, we show a global FIR-normalized CO SLED, constructed from the products of the inversion of the best-fit FIR/CO normalization parameter for each transition ($\bar{A}$, which is equal to $L_{\mathrm{FIR}}/L'_{\mathrm{CO}}$, given in Table \ref{Tab01}) $\times{J_u}^2$ (where ${J_u}^2$ is included to make the unit same as integrated line flux $Jy\:km\:s^{-1}$). 
Error bars are the dispersions of CO--FIR correlations (the $\sigma$ shown in Fig.\ref{Fig01}). 

This CO SLED represents the average CO excitation conditions in local galaxies, and reveals at least two components: a low-excitation component peaking around $J_{u}\sim3-4$, and a higher-excitation component peaking around $J_{u}\sim8$. 

We used RADEX \citep{vanderTak2007} to construct two-component LVG models and performed least-$\chi^2$ fitting to search for the best-fits of gas kinetic temperature $T_{\mathrm{kin}}$ and density $n_{\mathrm{H}_2}$, 
deriving $T_{kin}\sim90_{-40}^{+100}\;K$, $\log\,n_{\mathrm{H}_2}\sim3.0_{-0.3}^{+0.3}\;cm^{-3}$ for the lower-excitation component, and $T_{kin}\gtrsim200\;K$, $\log\,n_{\mathrm{H}_2}\sim4.1_{-0.2}^{+0.2}\;cm^{-3}$ for the higher-excitation component. 

In Fig.\ref{Fig04}, we also overlay the FIR-normalized CO SLEDs of a selection of galaxies that have at least six CO detections and span a wide $L_{\mathrm{FIR}}$ range. 
(U)LIRGs/starbursts have stronger high-{\it{J}} CO excitations and their normalized SLEDs are flatter at $J_{u}\gtrsim6$, whereas normal SFGs with weaker excitations show a bump at $J_{u}\sim4$ in their normalized SLEDs. 
Surprisingly, the largest difference between normalized SLEDs of (U)LIRGs and SFGs is seen not at a high-{\it{J}} (i.e. $J_{u}\sim10$) but at a low-$J$ (i.e. $J_{u}\sim4$), where the CO/FIR ratios vary within $\sim1.5\,dex$. 
Since the low-excitation component dominates the low-$J$ part of SLED, the large variation at a low-$J$ indicates that low-excitation gas is less correlated with FIR, while the high-excitation component is intrinsically better correlated with FIR \citep[see also][]{Lu2014}.

\subsection{The CO(5--4) -- FIR Correlation and Extension to High-{z}}
\label{Section33}

\begin{figure}
\epsscale{1.15}
\plotone{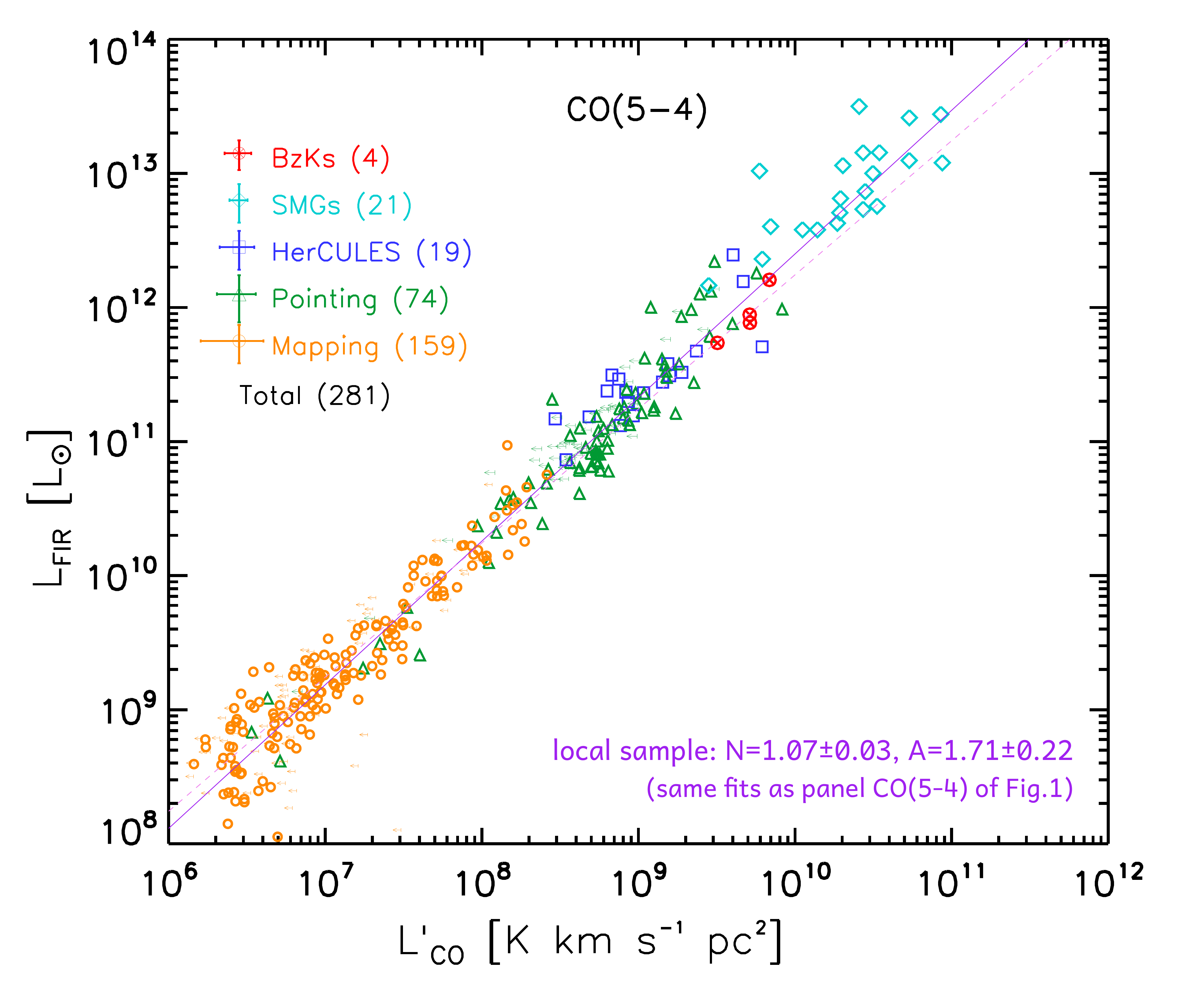}
\caption{
CO(5--4)--FIR correlation including high-{\it{z}} galaxies. 
The cyan diamonds are high-{\it{z}} galaxies from \citetalias{Carilli2013}. 
The crossed red circles are $z\sim1.5$ BzKs from \cite{Daddi2014}. 
The other data points and the best fit of the local sample are the same as the CO(5--4) panel of Fig.\ref{Fig01}. 
\label{One_CO_InBeam_IR_CO54}}
\end{figure}

To extend the high-{\it{J}} CO--FIR correlation toward higher-{\it{z}}, we use the extensive compilation of high-{\it{z}} CO data in \citetalias{Carilli2013} that contains 60 CO $J_{u}\ge4$ detections with literature $L_{\mathrm{IR}}$%
\footnote{The $L_{\mathrm{FIR}}$ for nine galaxies are derived from IR SEDs, a further 14 from FIR-radio correlation \citepalias{Carilli2013,Bothwell2013}, and the remaining 37 scaled from single-band $850\mu{m}$ or $1200\mu{m}$ \citepalias{Carilli2013,Greve2014}. Where required we converted $L_{\mathrm{TIR}}$ to $L_{\mathrm{FIR}}$ by a factor of $1/1.3$.}. 
They are the most extreme starbursts across all cosmic time, likely experiencing a short-term burst phase of SF resembling local ULIRGs, rather than the long-lasting mode in normal SFGs or the so-called ''main-sequence'' (MS) galaxies at high-{\it{z}} \citepalias[e.g.][]{Carilli2013}. 

CO $J_{u}\ge4$ observations toward high-{\it{z}} MS galaxies are still very rare: \cite{Daddi2014} presented the first CO(5--4) detections in a sample of 4 $z\sim1.5$ BzK-color-selected MS galaxies (BzKs). 
Thus combining all high-{\it{z}} $J_{u}\ge4$ data, CO(5--4) has the second-most detections but covers the most diverse galaxy types. 

In Fig.\ref{One_CO_InBeam_IR_CO54}, we show the CO(5--4)--FIR correlation, including high-{\it{z}} galaxies as an example to illustrate the correlation between dense gas and SFR, as CO(5--4) traces a hundred times denser molecular gas than CO(1--0). 
We show the linear fits from Fig.\ref{Fig01} only, as the inclusion of high-{\it{z}} SMGs and BzKs does not significantly change the results of the fit. 
The mean $\log\,(L_{\mathrm{FIR}}/L'_{\mathrm{CO}(5-4)})=2.27\pm0.07\;L_{\odot}\;(K\,km\,s^{-1}\,pc^{2})^{-1}$ is consistent with \cite{Daddi2014} considering a conversion factor of 1/1.3 from TIR to FIR. 
BzK galaxies fall on the local linear best-fit correlation (dashed line), 
whereas high-{\it{z}} SMGs are offset above by $\sim0.28\,dex$. 
Interestingly, using HCN(1-0) as a dense gas tracer, \cite{Gao2007} found higher FIR/HCN ratios in high-{\it{z}} galaxies (mostly QSOs/AGNs with only 5 HCN detections) than those of local ULIRGs/SFGs. 

Given the large uncertainties in $L_{\mathrm{FIR}}$ for a large fraction of SMGs 
\footnote{For example, $850\mu{m}$-derived $L_{\mathrm{FIR}}$ (using \citetalias{Carilli2013} conversion factor) are $\sim77-200\%$ of radio-derived $L_{\mathrm{FIR}}$ in \cite{Bothwell2013}. Lensing corrections also introduce uncertainties.} compared to BzKs (from SED fitting), the amount of excess in FIR/CO(5--4) (i.e. dense gas SF efficiency) in SMGs should still be treated with caution. 
And this excess is not significant enough to break down a linear form of dense gas versus SF relation, as these starbursts are the most extreme systems at all redshifts and as such do not represent the dominant mode of SF observed in MS galaxies (e.g. BzKs). 
Thus, better FIR measurements, e.g. via SEDs, and additional high-{\it{J}} CO  observations in normal MS galaxies are needed before making solid conclusions at high-{\it{z}}. 

\subsection{The CO(10--9) -- FIR Correlation Extending to Galactic Young Stellar Objects (YSOs)}
\label{Section34}

\begin{figure}
\epsscale{1.15}
\plotone{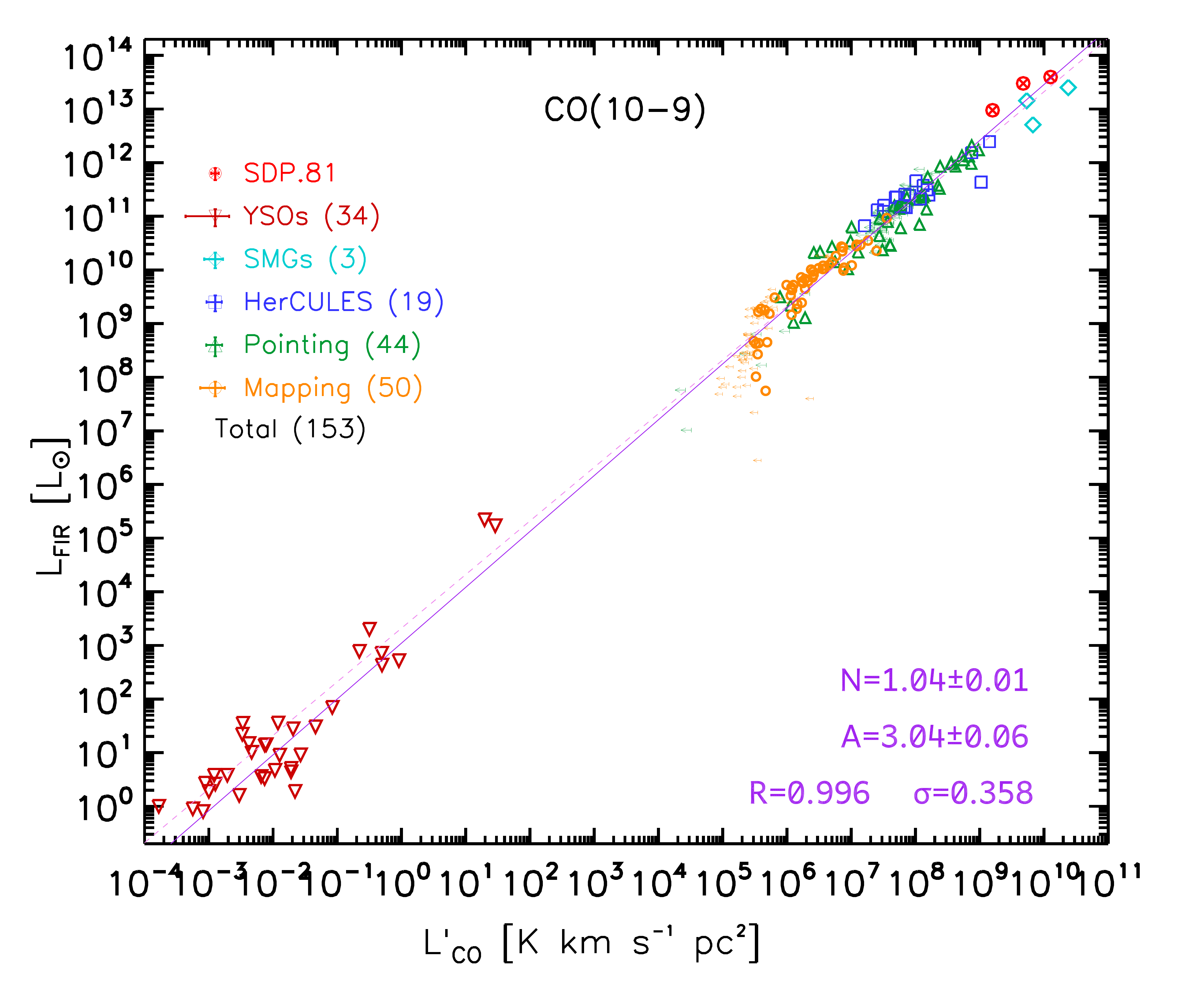}
\caption{
CO(10--9)--FIR correlation, including galactic YSOs/protostars \citep[][red triangles]{SanJoseGarcia2013}, high-{\it{z}} SMGs/QSOs \citepalias[][cyan diamonds]{Carilli2013}, and strongly lensed SMG SDP.81 \citep[global and two components;][red crossed circles]{ALMAPartnership2015}. 
The 3 data points of SDP.81 are: the west component, the east component, and global, from left to right, respectively. 
The solid line is a free-slope fit for all sources. 
The dashed line is a fixed-slope $N=1$ fit, indicating a mean  $\log\,(L_{\mathrm{FIR}}/L'_{\mathrm{CO}(10-9)})=3.30\pm0.09$.
\label{One_CO_InBeam_IR_SJ13}}
\end{figure}

In Fig.\ref{One_CO_InBeam_IR_SJ13} we show the correlation between an even denser/warmer gas tracer CO(10--9) and $L_{\mathrm{FIR}}$, which has the largest number of observations of sources spanning high-{\it{z}} galaxies (from \citetalias{Carilli2013} and \cite{ALMAPartnership2015}) to galactic YSOs/protostars (from \cite{SanJoseGarcia2013}).

\cite{ALMAPartnership2015} presented spatially resolved ALMA CO(10--9) in a strongly lensed SMG SDP.81 at $z=3.042$, with demagnified $L_{\mathrm{TIR}}=5.1\times10^{12}\,L_{\odot}$, comparable to the BzKs. We show the global SDP.81 as well as its east and west components in Fig.\ref{One_CO_InBeam_IR_SJ13}. 

The best fit for all sources is
$N=1.04\pm0.01$ and 
$A=3.04\pm0.06$ 
\citep[see also][who obtained the same $N=1.04$ with only six galaxies]{SanJoseGarcia2013}; considering YSOs/protostars alone we get $N=1.26\pm0.09$, which is likely biased by the limited number of bright sources ($L_{bol}\approx10^5\,L_{\odot}$). 
A fixed-slope $N=1$ fit indicates that a mean  $\log\,(L_{\mathrm{FIR}}/L'_{\mathrm{CO}(10-9)})=3.30\pm0.09\;L_{\odot}\,(K\,km\,s^{-1}pc^{2})^{-1}$ is valid for all sources within $\sigma\sim0.36\,dex$. 

\subsection{Conclusions}
\label{Section35}
We use {\it{Herschel}} SPIRE FTS observations of 167 local galaxies, including mapping data, to determine $L'_{CO}$ $J=4-3$ to $12-11$, and derive corresponding beam-matched $L_{FIR,b}$ for each CO line using PACS photometry. 
We find that these CO--FIR correlations are all linear, and that the non-linear result reported in \citetalias{Greve2014} can be attributed to the comparatively small number of galaxies in their sample. 
The overall linearity suggests that a universal CO/FIR SLED exists among these galaxies, which further reveals two excitation states of gas. The high-excitation CO SLED peaks at $J\sim7$, where the transitions are the best tracers of SFR \citep[e.g.][]{Lu2014}. 

At high-{\it{z}}, however, SMGs/QSOs have elevated FIR/CO(5--4) ratios relative to those seen in local galaxies and contemporary BzKs: this is because they are the most extreme types of galaxies and hence are not representative of the typical SFG population. 

A tight, linear correlation between CO(10--9)--FIR is shown to hold for the majority of local galaxies, resolved sub-kpc regions, Galactic YSOs, and also high-{\it{z}} galaxies. 
These results strongly support a fundamental linear relationship between dense gas and the SFR, and also provide the local benchmark for probing the gas and SF in more types of galaxies (e.g. MS galaxies) at high-{\it{z}}. 

\acknowledgments
This work is supported by NSFC \#11173059, \#11390373, \#11420101002, and CAS \#XDB09000000. 
%
%
D. L. gratefully thanks T. Greve, Z. Zhang, P. Papadopoulos, S. Madden, and R. Wu for constructive discussions, and K. Okumura and B. Altieri for helpful instructions on PACS. 

\end{document}